\documentclass[sigconf]{acmart}
\usepackage{algorithm}
\usepackage{algpseudocode}
\usepackage{subfiles}
\usepackage{multirow}
\usepackage[utf8]{inputenc}
\AtBeginDocument{%
  }

\setcopyright{acmlicensed}
\copyrightyear{2025}
\acmYear{2025}
\acmDOI{XXXXXXX.XXXXXXX}

\begin{document}

\title{On Improving PPG-Based Sleep Staging: A Pilot Study}
\author{Jiawei Wang}
\affiliation{
  \institution{University of Warwick}
  \city{Coventry}
  \country{UK}
}
\email{davy.wang@warwick.ac.uk}

\author{Yu Guan}
\affiliation{
  \institution{University of Warwick}
  \city{Coventry}
  \country{UK}
}
\email{yu.guan@warwick.ac.uk}

\author{Chen Chen}
\affiliation{
  \institution{Fudan University}
  \city{Shanghai}
  \country{China}
}
\email{chenchen_fd@fudan.edu.cn}

\author{Ligang Zhou}
\affiliation{
  \institution{Fudan University}
  \city{Shanghai}
  \country{China}
}
\email{21110720086@m.fudan.edu.cn}

\author{Laurence T. Yang}
\affiliation{
  \institution{Zhengzhou University}
  \city{Zhengzhou}
  \country{China}
}
\email{ltyang@gmail.com}

\author{Sai Gu}
\affiliation{
  \institution{University of Warwick}
  \city{Coventry}
  \country{UK}
}
\email{sai.gu@warwick.ac.uk}

\renewcommand{\shortauthors}{Wang et al.}

\begin{abstract}
Sleep monitoring through accessible wearable technology is crucial to improving well-being in ubiquitous computing. 
Although photoplethysmography (PPG) sensors are widely adopted in consumer devices, achieving consistently reliable sleep staging using PPG alone remains a non-trivial challenge. 
In this work, we explore multiple strategies 
to enhance the performance of PPG-based sleep staging.
Specifically, we compare conventional single-stream model with dual-stream cross-attention strategies, based on which complementary information can be learned via PPG and PPG-derived modalities such as augmented PPG or synthetic ECG.
To study the effectiveness of the aforementioned approaches in four-stage sleep monitoring task, 
we conducted experiments on the world's largest sleep staging dataset, i.e., the Multi-Ethnic Study of Atherosclerosis (MESA). 
We found that substantial performance gain can be achieved by combining PPG and its auxiliary information under the dual-stream cross-attention architecture. 
Source code of this project can be found at: https://github.com/DavyWJW/sleep-staging-models.
\end{abstract}

\ccsdesc[500]{Applied computing~Health informatics}

\keywords{Sleep stage classification,  Cross-attention, Diffusion models, Wearable sensors}

\maketitle
\section{Introduction}
Sleep plays a vital role in human physical and mental health, and disturbed sleep has been associated with a wide range of chronic diseases, including chronic insomnia, depression, and cognitive decline \cite{ting2005disorders}. 
According to the American Academy of Sleep Medicine (AASM) \cite{berry2012aasm}, sleep is divided into wake, Rapid Eye Movement (REM) and three non-REM stages: N1, N2, and N3. Accurate identification of these stages is essential for diagnosing sleep disorders and evaluating sleep quality \cite{JBHI_25}.

The gold standard for sleep staging is polysomnography (PSG) \cite{rundo2019polysomnography}, which requires the analysis of multiple biosignals including electroencephalography (EEG), electrocardiography (ECG), electromyography (EMG), and electrooculography (EOG), etc.  
However, PSG is expensive, burdensome, and therefore difficult to scale for everyday use. 
This motivates research into more accessible methods of sleep monitoring, such as using wearable sensors \cite{Zhai20, sleepAI_review2020, sleep_wearable}.
Among wearable sensing modalities, photoplethysmography (PPG) has gained widespread adoption due to its low cost and ease of deployment \cite{habib2022performance}. 
Despite the advantages of PPG for lightweight sleep monitoring, there are opportunities to enhance its representational capabilities through innovative computational approaches \cite{kotzen2022sleepppg}. 

Recent advances have identified two particularly promising pathways among these opportunities. 
First, leveraging generative AI may enrich the input modalities, yielding improved results \cite{lan2023performer}. 
In \cite{shome2024region}, Region-Disentangled Diffusion Model (RDDM) was proposed for 
high-fidelity PPG-to-ECG synthesis.
The generated ECG was further used as auxiliary modality to many downstream tasks such as blood pressure estimation, stress detection with improved results \cite{shome2024region}. 
This raises the intriguing question: can synthetic ECG also be used to enhance the performance of sleep staging? 
Second, 
multi-stream processing and attention mechanism demonstrated their great potential in physiological signal analysis and modeling
\cite{eldele2021attention,supratak2017deepsleepnet}. 
Through attention-based deep learning approaches, promising results can be achieved for EEG-based sleep monitoring \cite{eldele2021attention}. 
In \cite{supratak2017deepsleepnet}, the dual-stream CNN with different filter sizes was used to extract 
complementary features from single-channel EEG to improve the performance in sleep monitoring. 

Motivated by these works, in this paper we study the effectiveness of the dual-stream attention structures and different PPG-derived modalities for 4-stage sleep monitoring, i.e., to classy Wake, REM, Light (N1+N2), and Deep Sleep.  
Specifically, three strategies will be investigated based on different auxiliary modalities, such as Augmented PPG, synthetic ECG, and Real ECG. 
Out of them,  synthetic ECG can be generated by generative models such as the pre-trained RDDM \cite{shome2024region}, while Real ECG can be acquired by using ECG sensor. 
All models are trained and validated on the Multi-Ethnic Study of Atherosclerosis (MESA) dataset \cite{chung2023multi} to ensure fair comparison and identify effective strategies for PPG-based sleep monitoring.

\section{Related Work}
Architectural design plays a key role in learning effective representations for sleep staging. Two-stream architectures, originally used in computer vision \cite{simonyan2014two}, process different input views in parallel and enable bidirectional information exchange between streams. 
In sleep analysis, cross-modal attention has been successfully applied to fuse biosignals such as EEG \cite{eldele2021attention, yu2025casleepnet}, where each modality can query and integrate complementary information from others. 
However, these approaches have not been explored for PPG-based sleep staging. 
Given the distinct characteristics of PPG signal compared to EEG (e.g., pulse activity vs. neural activity), it remains unclear whether dual-stream processing with cross-attention mechanism can effectively enhance PPG-based sleep classification. This motivates our investigation of adapting these architectural innovations to PPG-based sleep monitoring tasks. 

Generative AI has shown remarkable success in domains such as text \cite{brown2020language} and image generation \cite{ramesh2022hierarchical}.
Recently, generative models have been applied to 
biosignal generation and augmentation, particularly in synthesizing ECG-like waveforms from PPG \cite{sarkar2021cardiogan, lan2023performer}. Building on the denoising diffusion probabilistic model (DDPM) \cite{ho2020denoising}, RDDM was proposed in \cite{shome2024region}, which enables generation of high-fidelity, morphology-aware ECG signals. Motivated by these approaches, in this paper we investigate whether synthetic ECG can improve the performance of PPG-based sleep staging. 

\section{Methodology}
In this section, we introduce two types of sleep staging models, namely, single-stream model and dual-stream model. 
Based on the dual-stream structure, different modalities can be combined to learn complementary information for improved performance. 

\subsection{Single-stream Model}
\begin{figure}[!htbp]
    \centering
    \includegraphics[width=0.35\textwidth]{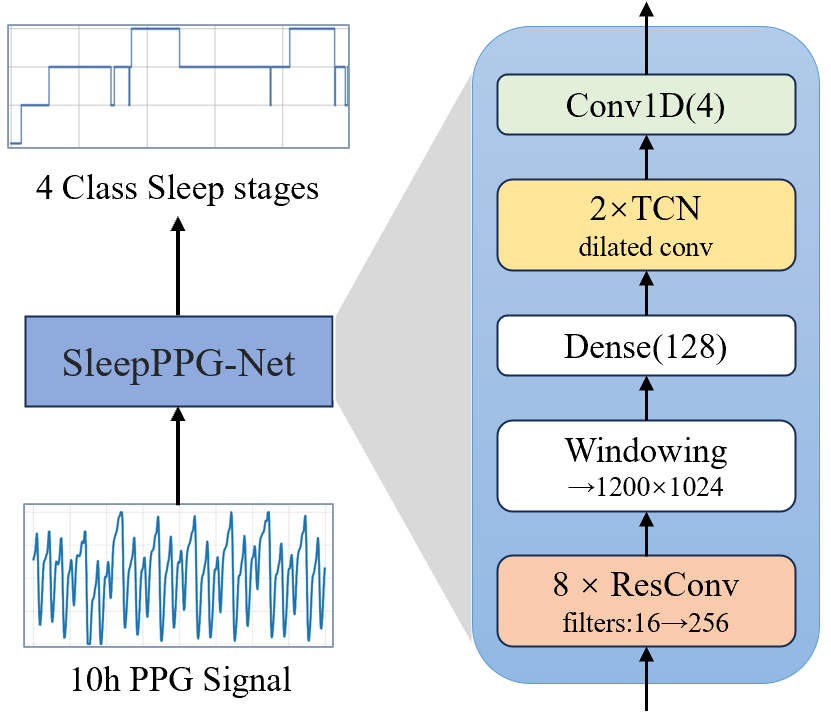}
    \caption{Single-stream sleep staging model (i.e., SleepPPG-Net\cite{kotzen2022sleepppg})}
    \label{fig:single}
\end{figure}

For single-stream model, we use the classical SleepPPG-Net baseline \cite{kotzen2022sleepppg}, which
processes 10-hour PPG recordings through a sequence of residual convolutional blocks followed by temporal convolution layers, enabling end-to-end 4-class sleep stage prediction. 
This single-stream model provides a 
robust foundation for comparing enhancement strategies. 
The architecture of this single-stream model can be found in Fig.~\ref{fig:single}.

\begin{figure}[!htbp]
    \centering
    \includegraphics[width=0.25\textwidth]{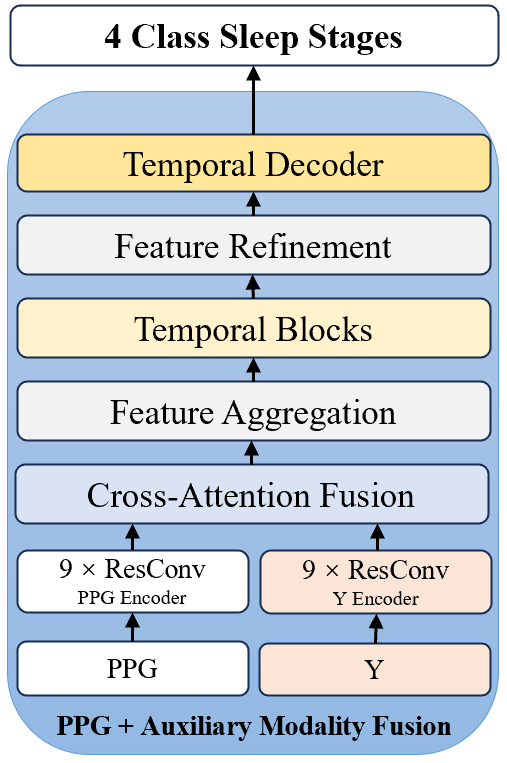}
    \caption{Dual-stream sleep staging model}
    \label{fig:dual}
\end{figure}

\subsection{Dual-stream Model}

The dual-stream model extends the single-stream architecture to incorporate PPG and an auxiliary modality $Y$, where $Y$ can be PPG-derived modalities such as synthetic ECG, augmented PPG, or other relevant physiological signals such as ECG. 
As shown in Fig. ~\ref{fig:dual}, the design of this architecture is modality-independent and intended to support general multi-modality fusion.
It includes 
\textbf{dual-stream encoders} (9 ResConvBlocks each) to extract features from both input signals in parallel, followed by \textbf{cross-attention fusion} to integrate the two streams via bidirectional cross-attention modules. Based on this architecture, we study three dual-stream cases:

\begin{itemize}

\item \textbf{PPG + Augmented PPG:} We create augmented PPG (i.e., $Y$) by adding controlled noise components (Gaussian noise, baseline drift, motion artifacts). 
Based on this simple PPG-derived modality, we aim to explore whether cross-attention can extract complementary information from signal variations without new physiological information.

\item \textbf{PPG + Synthetic ECG:} We use the pre-trained RDDM \cite{shome2024region} to synthesize ECG signals given the PPG inputs.
The synthesized ECG is used as the auxiliary modality $Y$ for the dual-stream sleep staging model. 
This approach evaluates whether synthetic physiological signals (i.e., synthetic ECG) can enhance sleep staging performance.

\item \textbf{PPG + Real ECG:} 
This variant incorporates Real ECG and PPG signals, which are processed using the dual-stream sleep staging model. 
This configuration serves as the upper limit of performance of PPG+ Synthetic ECG. 


\end{itemize}

\textbf{Remark:}
To enable interaction between dual streams, we adopt a bidirectional cross-attention mechanism between PPG and auxiliary inputs. Each modality selectively attends to relevant temporal patterns in the other stream using multi-head attention, followed by layer normalization and feed-forward networks. This design encourages effective integration of complementary information and supports alignment in both time and feature domains.
More details can be found in Appendix A.

\section{Experimental Evaluation}

\subsection{Dataset and Experimental Settings}
The MESA Sleep Study represents one of the largest and most diverse polysomnography datasets available for sleep research \cite{chung2023multi}. The dataset comprises overnight recordings from 2,056 diverse ethnic participants aged 54-93 years. Each recording includes synchronized PPG signals from finger pulse oximetry, Lead-II ECG at 256 Hz, and expert-annotated sleep stages following AASM guidelines \cite{chung2023multi}. The test set follows \cite{kotzen2022sleepppg}, comprising 204 predefined subjects, with the remaining 1,852 subjects split into training (1,481 subjects) and validation (371 subjects) sets. Each subject contributes a whole night of 10-hour sleep recording, divided into 1,200 30-second epochs. Sleep stage predictions are made on a per-epoch basis, with classification performance measured using accuracy and Cohen's kappa  score ($\kappa$).

\begin{table}[ht]
\centering
\begin{tabular}{lcc}
\toprule
\multirow{2}{*}{Strategy} & \multicolumn{2}{c}{MESA} \\
\cmidrule(r){2-3}
& $\kappa$ & Accuracy \\
\midrule
PPG & 0.675 & 0.783 \\
Augmented PPG& 0.630 & 0.754 \\
\midrule
PPG + Augmented PPG & \textbf{0.745} & \textbf{0.833} \\
PPG + Synthetic ECG & 0.723 & 0.820 \\
\midrule
PPG + Real ECG & 0.739 & 0.829 \\

\bottomrule
\end{tabular}
\caption{Performance (in accuracy and Cohen's kappa $\kappa$) of single-stream and dual-stream sleep staging models.}
\label{tab:performance_comparison}
\end{table}


\subsection{Preprocessing}
Following the procedure in \cite{kotzen2022sleepppg}, PPG signals were preprocessed as follows: 
they were low-pass filtered using a Chebyshev Type II filter (8 Hz cutoff), resampled to 34.133 Hz, outlier-clipped at 3 standard deviations, and standardized using z-score normalization. 

For ECG signals, we followed the preprocessing procedure in \cite{buendia2012high}. 
ECG signals were band-pass filtered (0.5–40 Hz) using a 4th-order Butterworth filter, resampled, and normalized to zero mean and unit variance.

For synthetic ECG generation, we employed the pre-trained RDDM model \cite{shome2024region}. 
The input PPG signals were upsampled to 128 Hz, segmented into overlapping 4-second windows, processed through RDDM, and the generated ECG segments were reconstructed and downsampled back to 34.133 Hz.

For augmented PPG, we created it by adding controlled noise components to the pre-processed PPG. 
Specifically, we added Gaussian noise ($\sigma = 0.1$), baseline drift (amplitude = 0.1, frequency = 0.1 Hz), and motion artifacts (spike probability = 0.01, amplitude = 0.5). 
This creates signal variations while preserving the underlying physiological patterns.

\subsection{Implementation Details}
All models are implemented in PyTorch and trained using the AdamW optimizer with a learning rate of $1 \times 10^{-4}$, weight decay of $1 \times 10^{-5}$, and batch size 2. 
Training continues for a maximum of 50 epochs with early stopping based on validation Cohen's kappa. 
Each model processes 10-hour continuous recordings (1,228,800 samples) organized into 1,200 30-second windows.

\textbf{Single-stream model:} Following \cite{kotzen2022sleepppg}, single-stream model employs 8 ResConvBlocks for feature extraction, followed by a temporal windowing layer, time-distributed dense layer (128 units), and 2 stacked TCN blocks for temporal modeling.

\textbf{Dual-stream model:} dual-stream model employs dual-stream encoders with 9 ResConvBlocks per stream, followed by 3 bidirectional cross-attention modules (feature dimension = 256, 8 heads) for inter-modal fusion. After fusion, the architecture uses identical temporal modeling as the single-stream counterpart. 

\subsection{Results and Analysis}

Table~\ref{tab:performance_comparison} reports Cohen's kappa ($\kappa$) and accuracy with respect to the different strategies on MESA test set.
We can observe that PPG + Augmented PPG achieves the highest performance with an accuracy of 83.3\% and kappa 0.745, improving accuracy by 5\% over the PPG-only single-stream baseline (78.3\%), suggesting the effectiveness of the dual-stream cross-attention architecture.
However, under the same dual-stream structure, the PPG + synthetic ECG strategy shows inferior performance.
One reason can be the inconsistent quality of synthetic ECG --- generated by RDDM\cite{shome2024region}, which was trained based on non-sleep PPG-ECG pairs. 

It is interesting to see that PPG + Augmented PPG strategy has a performance comparable to the PPG + Real ECG counterpart.
This observation indicates the effectiveness of dual stream cross-attention architecture in extracting complementary information within a single PPG modality (i.e., on par with PPG+ECG modalities).  

\begin{figure}[!htbp]
    \centering
    \includegraphics[width=0.5\textwidth]{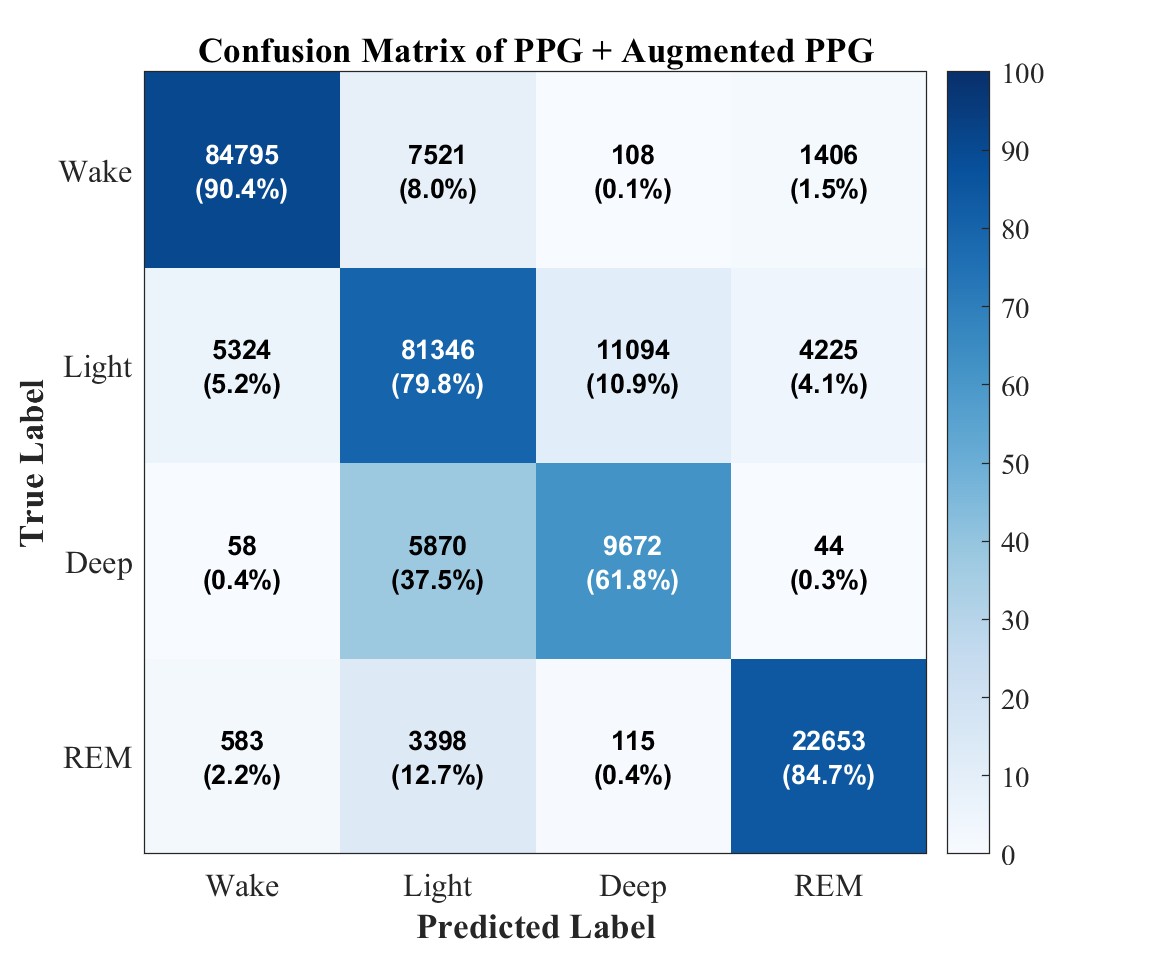}
    \caption{Confusion matrix (PPG + Augmented PPG strategy)}
    \label{fig:p4}
\end{figure}

To understand better the dual-stream structure, we also report result of augmented PPG under the single-stream structure in Table~\ref{tab:performance_comparison}. 
We can see although augmented PPG alone performs worse than PPG (75.4\% vs 78.3\% in accuracy), substantial performance gain can be achieved by combining them via the two-stream sleep staging model, 
confirming its capacity to extract discriminative features from signal variations.






Fig.~\ref{fig:p4} presents the confusion matrix for the best model (i.e.,  PPG + Augmented PPG strategy) on the MESA test data. 
We can see this approach achieves high accuracy for classifying Wake (90.4\%), Light (79.8\%), and REM (84.7\%) classes. 
Deep sleep remains the most challenging stage to be recognized, with an accuracy of $61.8\%$, and 37.5\% of them were misclassified as Light sleep. The relatively lower performance on identifying Deep sleep (N3) is consistent with previous studies \cite{kotzen2022sleepppg, attia2024sleepppg} and reflects the inherent challenge of detecting this stage using peripheral signals like PPG.



\textbf{Remark:} 
by extracting complementary information, the dual-stream sleep staging strategies can achieve much higher performance than single-stream solutions.  
While the cross-modal model leverages synthetic ECG (or Real ECG) to enhance the recognition accuracy,
PPG and ECG are highly correlated and may have some upper bound in terms of multimodal fusion \cite{shome2024region}. 
In contrast, augmented PPG offers multiple advantages: it is easy to generate, and it eliminates the needs for additional device (i.e., ECG sensor), or external generative AI models (i.e., RDDM \cite{shome2024region}), making it a practical solution for improving the performance of PPG-based sleep staging.


\section{Conclusion}

In this pilot study, we investigated multiple strategies to improve the performance of PPG-based sleep staging. 
We studied both single-stream and dual-stream sleep staging models, and we found under the dual-stream cross-attention fusion structure, substantial performance gain can be achieved by combining PPG and PPG-derived signals such as augmented PPG, and synthetic ECG. 
In the future, we will study other auxiliary modalities such as EEG, explore 
modality-specific architectures for better multi-modal fusion, and also evaluate our models on more sleep staging datasets.

\appendix

\section*{Appendix}
\addcontentsline{toc}{section}{Appendix}

\renewcommand{\thesection}{\Alph{section}.}

\section{Cross-Attention Fusion}

Bidirectional cross-attention is applied once features $F_{PPG}$ and $F_Y$ 
(where $Y$ can be real ECG, synthetic ECG, or augmented PPG) are extracted by 
the dual encoders. The mechanism enables each stream to selectively attend 
to relevant temporal patterns in the other stream through $N_f=3$ fusion 
blocks applied sequentially.

For each direction, queries ($Q$) come from the target stream, keys/values 
($K,V$) from the source stream:
\[
F^{out}=\operatorname{LN}\bigl(F+\operatorname{MHA}(Q,K,V)\bigr),\quad 
Q=W_q F,\; K=W_k F',\; V=W_v F'
\]
where $\operatorname{LN}$ denotes layer normalization, $\operatorname{MHA}$ represents multi-head attention, and $W_q$, $W_k$, $W_v$ are learnable projection matrices.

After cross-attention, a feed-forward network (FFN) produces refined features. 
We use $H=8$ heads and feature dimension=256. Following \cite{simonyan2014two}, adaptive modality weighting is applied 
before final aggregation:
\[
\hat{w}_{PPG} = \frac{\sigma(\mathrm{MLP}(\mathrm{GAP}(F_{PPG})))}
{\sigma(\mathrm{MLP}(\mathrm{GAP}(F_{PPG}))) + \sigma(\mathrm{MLP}(\mathrm{GAP}(F_{Y}))) + \epsilon}
\]
where $\mathrm{GAP}$ denotes global average pooling, $\mathrm{MLP}$ is a multi-layer perceptron, $\sigma$ is the sigmoid activation function, and $\epsilon$ is a constant for numerical stability.
\vspace{4pt}




\bibliographystyle{unsrt}
\bibliography{references}

\end{document}